# Exploring players' experience of humor and snark in a grade 3-6 history practices game

David J. Gagnon, University of Wisconsin
Ryan S. Baker, University of Pennsylvania
Sarah Gagnon, University of Wisconsin
Luke Swanson, University of Wisconsin
Nick Spevacek, University of Wisconsin
Juliana Andres, University of Pennsylvania
Erik Harpstead, Carnegie Mellon University
Jennifer Scianna, University of Wisconsin
Stefan Slater, University of Pennsylvania
Maria O.C.Z. San Pedro, University of Pennsylvania

**Abstract:** In this paper we use an existing history learning game with an active audience as a research platform for exploring how humor and "snarkiness" in the dialog script affect students' progression and attitudes about the game. We conducted a 2x2 randomized experiment with 11,804 anonymous 3rd-6th grade students. Using one-way ANOVA and Kruskall-Wallis tests, we find that changes to the script produced measurable results in the self-reported perceived humor of the game and the likeability of the player character. Different scripts did not produce significant differences in player completion of the game, or how much of the game was played. Perceived humor and enjoyment of the game and its main character contributed significantly to progress in the game, as did self-perceived reading skill.

**Introduction**

Video games have been repeatedly proven to be useful tools for learning across a variety of learning goals (e.g. Clark et al., 2016), and have been linked to increased motivation and interest in learning (National Research Council, 2011). Clark et al., in their 2016 metastudy exploring the effectiveness of games suggest that

> [Future] research on games and game-based learning should thus shift emphasis from proof-of concept-studies ("can games support learning?") and media-comparison analyses ("are games better or worse than other media for learning?") to value added comparisons and cognitive-consequences studies exploring how theoretically-driven design decisions influence learning outcomes for the broad diversity of learners within and beyond our classrooms.

One specific area of design is the construction of the game's narrative. Narrative has been previously studied as a means for sustaining engagement and supporting learning (Dickey, 2011). As an attribute of narrative, Dormann & Biddle (2006), explore humor in digital games, arguing that it contributes to enjoyment of the experience, as well as learning by sustaining emotional and cognitive engagement, as well as stimulating social presence. Peng et al. (2010) also provide a useful insight into the role of game narrative, demonstrating that players of the game Darfur is Dying engage in identification (Cohen, 2001) and role-taking (Davis et al., 1996) with the game's main character.

*Research Agenda*

In this paper, we study how humor and a "snarky" writing style for the player character relate to different player attitudes about the game and progress through the game. Given that the humorous and snarky elements are experienced through reading, we consider the age and self-perceived reading skill in relation to how far players progress in the game. We also explore how their attitudes toward the game and toward history affect their progression in the game.

Three research questions drive this investigation:

> **Research Question 1:** How does snark and humor in the game script influence players' perceptions of the game and the domain of history?
>
> **Research Question 2:** Do snark or humor in the script cause players to play longer and complete more of the game?
>
> **Research Question 3:** How do other factors such as self-perceived reading skill and attitudes toward the game affect the amount of the game completed?

*Jo Wilder and the Capitol Case*
To conduct this experiment we utilize *Jo Wilder and the Capitol Case* (2019), a point and click adventure game designed for grade 3-5 students to learn historical practices in context of investigating the women's suffrage movement and the creation of earth day. Distributed by PBS learning media, PBS Wisconsin and BrainPOP, the game has been played over 250,000 times since release in 2019.

**Methods**

*Participants*

Data was collected from 11,804 anonymous players who self-identified as grade 3-6 students (typically aged 8-11 years-old), between September 1, 2021, and November 30, 2021. Analytics indicate that the vast majority of these sessions (>95%) were from the United States, and >80% of these sessions took place during the local time of 8am to 4pm on non-holiday weekdays, leading us to believe that the majority of these sessions took place in formal learning contexts.

*Procedure*

***Different versions of the game script were developed and randomly assigned to players***
Four scripts were created to explore how changes to the game's writing alone (no game mechanics, level design, or changes of any other sort were made) would potentially affect the ways the players experienced and progressed in the game. In total, the original version of the game has 597 lines in the script. As part of this re-writing effort, the original game writer created four versions of the game script that removed two key components of the game's writing, a "snarky" protagonist and humorous elements from the entire cast of characters. This resulted in four versions of the script:

**Snark + Humor (Original)**. This was the original script that includes a "snarky" protagonist, Jo Wilder, and a variety of jokes, gags and puns throughout the story. Jo Wilder is not a "model student." Jo is not an earnest lover of history, excited by every artifact and topic of the game. Quite the contrary, she would rather go outside and play than do her homework, and she thinks history is "boring." The script also was written to include numerous small jokes, word plays and gags. Jo does not undergo a metamorphosis during the story, wrapping up the game with a newfound interest in the practices of history. These decisions were made while envisioning a typical 4th grader who was assigned the game to play, who doesn't necessarily like history, and would be respected for their preferences, not made to identify with a history-loving player character.

In this script Jo Wilder believes that meetings, historical artifacts and school are boring and says so often. She dislikes another character, Wells, because she believes that he treats her grandfather disrespectfully and later because she believes that he stole her pet badger.

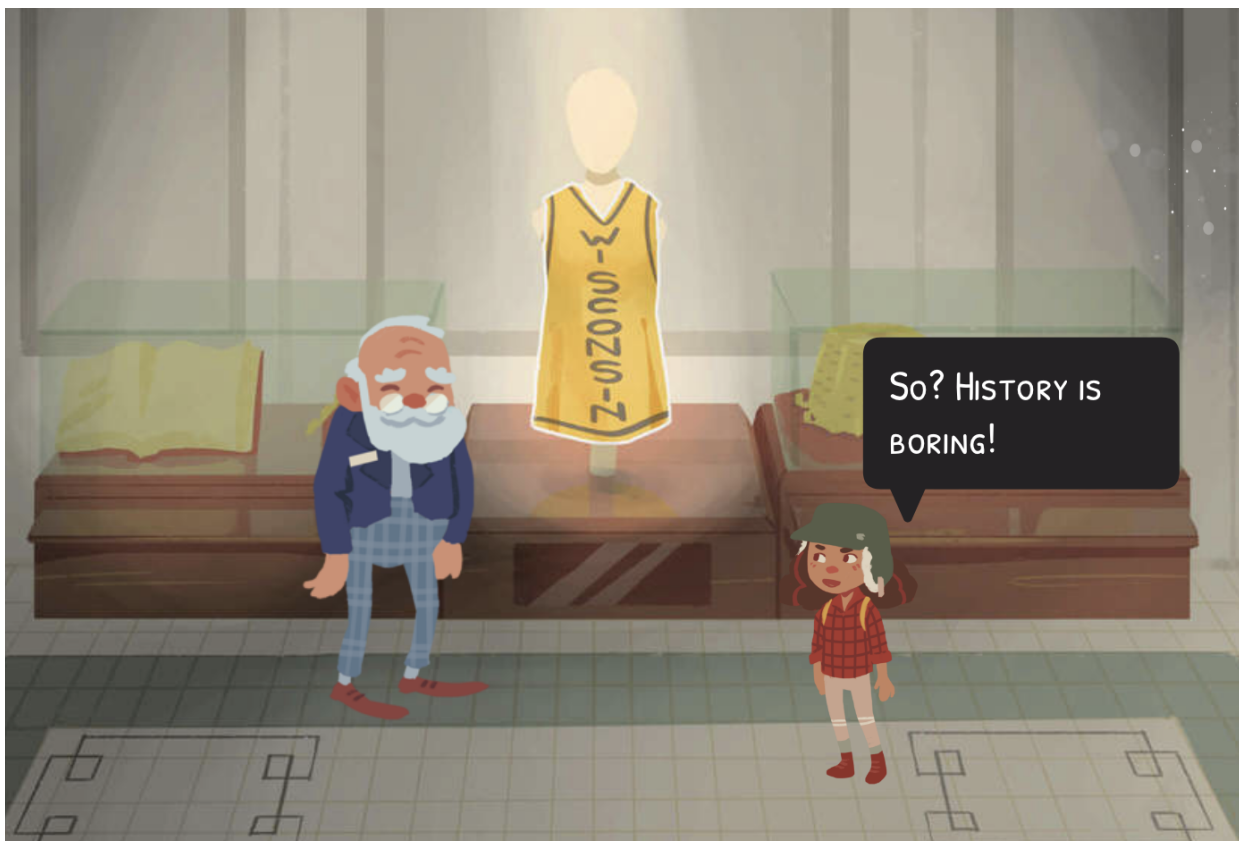

**Figure 1:** Jo's attitude in the original (Snark+Humor) condition.

**Humor, no Snark**. In this version, the Jo Wilder character was rewritten to seem "earnest." She acts as a model student, is generally interested in everything, likes school and is respectful to all adults. An example of this edit can be seen in an interaction between Gramps and Jo about her homework (Table 1). Humorous script elements not related to Jo's snarkiness were preserved. In this version, 71 lines, roughly 12%, of the normal script was changed or deleted.

**Snark, no Humor**. In the No Humor script, all characters' dialog was inspected for intentionally humorous elements such as jokes, wordplay and silly vocabulary. These lines were rewritten to preserve any necessary meaning while removing the humor, or deleted. For example, in the original version the Boss says, "Wells got in trouble for littering at the Wildlife Center. PLEASE tell me you're doing better than he is." In the edited version, this interaction is removed, and the boss simply says, "The exhibit opens tomorrow. How's the investigation going?" Another example is one of the many lines from Gramps that uses silly vocabulary. In the original version, Gramps says, "Look at that! It's the bee's knees!" which is replaced by the more neutral "What a fascinating artifact!" Jo's snarky lines were not modified. As a result, 147 lines, roughly 25% of the script, was edited or deleted.

**Dry.** The Dry script attempted to remove both snark and humorous elements, leaving the script as bare as possible of these elements while retaining meaning. 197 lines, approximately 33% of the script, were edited or deleted.

Logic was added to the game code to randomly assign a version of the script to each player upon game start. This means that within a given classroom implementation of the game, each student was equally likely to experience any of the four scripts and students in the same classroom played different versions of the game.

**Table 1**
Comparing the Original script to the No Snark and Dry scripts.

| Original Script | No Snark and Dry Scripts |
|---|---|
| **Gramps:** Your teacher said you missed 7 assignments in a row! <br> **Jo:** Well, I did SOME of those. I just couldn't find them! <br> **Gramps**: Did you do all of them? <br> **Jo:** No… because history is boring! <br> **Gramps**: I suppose historians are boring, too? <br> **Jo**: No way, Gramps. You're the best! | [Lines Deleted] <br><br> **Jo:** It's already all done! <br><br> [Lines Deleted] <br><br> **Jo:** Plus, my teacher said I could help you out for extra credit! |

*Instruments*
A total of five questionnaires were added to the game to determine the players' grade level and self-assessed reading skill, their enjoyment of the game, their interest in history, Jo's likeability, and how funny they perceived the game to be (Table 2). Each questionnaire appears at the beginning of a chapter in the game, a total of five times. The first questionnaire asks about the grade level and reading level. The following questionnaires use one item from each of the four subscales.

To assess the player's enjoyment of the game, items from the EGameFlow measure (Fu et al., 2009) were adopted. For interest in history, items were adopted from the Three-Dimensions of Student Attitude Towards Science (TDATS; Zhang & Campbell, 2011). For Jo's likeability, items were adopted from the Reysen Likability Scale (Reysen, 2005). The humor score was adapted from the Multidimensional Sense of Humor Scale (MHS; Thorson & Powell, 1993). Language from the MHS was adjusted in two main ways: 1) to reflect language that would be appropriate for the age group of students using Jo Wilder and 2) to refer specifically to agents and characters within the context of the study. Responses for these items were 5-point Likert scales from disagree to agree. Responses were reverse coded (scaled from agree to disagree) for every other question, making it possible to filter out sessions where the learner was not seriously attempting to answer the questions.

Each subscale was tested using Cronbach's Alpha. Enjoyment (α=.62), history interest (α=.88), Jo's likeability (α=.74), and the game's humor (α=.80) were all found to be reliable. For analysis that use the subscales, equal weighting is given to all items in a subscale.

**Table 2**
Questionnaire items.

| | |
|---|---|
| Player grade and reading level | Q1: What grade are you in? (3rd, 4th, 5th, 6th, Other)<br>Q2: How well do you read in English? (Not a very good reader, An OK reader, Very good reader) |
| Enjoyment of the game | Q3: The game grabs my attention.<br>Q8: Time flies while I'm playing the game.<br>Q13: I forget what's around me while playing the game.<br>Q18: I feel emotionally involved in the game. |
| Interest in History | Q4: I like watching TV shows about history.<br>Q9: I like reading about history.<br>Q14: I like learning history very much.<br>Q15: I think learning history is fun. |
| Jo's likability | Q5: Jo is friendly.<br>Q10: I like Jo.<br>Q11: Jo is kind.<br>Q16: I can relate to Jo. |
| Game humor | Q6: I think the characters are funny.<br>Q7: The characters say things that make me laugh.<br>Q12: The characters say funny things.<br>Q17: The characters are entertaining. |

*Data collection, preprocessing and analysis*
The game was publicly deployed and data collection began in September, 2021. All game telemetry events, including the questionnaires and the randomly assigned script id, were recorded with an anonymous session identifier. Session features were engineered for session duration and maximum level achieved. These files were then imported into a Google Colaboratory notebook using pandas and matlab for processing and visualization.

Sessions were removed if they had an outlyingly large duration (chosen at >4000 seconds), if they didn't finish the first chapter and therefore never answered the second questionnaire, or if the player was not using the newest version of the game (v10). Since the game does not track individual player identities, sessions that used a save code were also excluded. This resulted in sessions from 11,804 players.

Statistical analyses were conducted to test the relationship between the presence of snark and humor in the game script (four script types) and the player's self-reported perception of the game (enjoyment of the game, Jo's likeability, game humor) and interest in History (RQ1). One-way ANOVA tests (or, where necessary, its non-parametric equivalent Kruskall-Wallis) were conducted to determine if differences in game scripts had any effect on the players' self-reported measures. The same type of statistical tests were also employed to determine if the presence of snark and humor in the game script had any relationship with the players' completion of the game and how much progress the player made in the game, operationalized as the maximum level they reached in the game (RQ2). And lastly, we analyzed whether players' completion of the game and maximum level reached could be explained by their perceptions of the game (enjoyment of the game, Jo's likeability, game humor) and their reading level (RQ3).

## Results and Discussion

Our final sample used in analysis consisted of 7,116 players. Table 3 shows the distribution of the self-report attitudinal measures -- all four self-report measures had mean values over 0 (Neutral on the scale used) for all conditions. Approximately 2/3 of the players reported an above average reading level (66.3%); 25.9% reported an average reading level, and the remainder of the respondents reported a below-average reading level. The majority of the players did not complete the game (74.3%) and the mean time spent playing the game was 20 minutes, 30 seconds (SD = 12 minutes, 3 seconds).

**Table 3**
Distribution of players' self-reported measures by condition. All scales -2 (lowest) to +2 (highest). Means and standard deviations shown. Conditions statistically significantly higher than at least one other condition for a given metric are denoted in boldface italic.

| Self-report measure | Original (Snark, Humor) | Humor (No Snark) | Snark (No Humor) | Dry (No Snark or Humor) |
|---|---|---|---|---|
| Enjoyment of the game | 1.29 (1.12) | 1.34 (1.06) | 1.28 (1.09) | 1.28 (1.11) |
| Interest in history | 0.12 (1.44) | 0.20 (1.43) | 0.15 (1.43) | 0.21 (1.44) |
| Jo's likeability | 0.92 (1.20) | ***1.29 (1.06)*** | 0.91 (1.16) | ***1.30 (1.03)*** |
| Game humor | ***0.20 (1.43)*** | ***0.22 (1.44)*** | 0.13 (1.43) | 0.06 (1.41) |

When analyzing whether the effects of game script type on player attitudes, only the perception of Jo's likability and game humor yielded statistically significant differences between scripts. There was a significant effect of game script type on player perception of Jo's likability for the four game script types, $\chi^2$ (df = 3)= 235.6, p<.0001, as well as on the player's perception of game humor for the four game script types, $F(3, 7112) = 4.86$, p = 0.002. There was not a statistically significant effect for enjoyment, $F(3, 7112) = 1.19$, p=0.31, or for interest in history, $F(3, 7112)=1.59$, p=0.19.

Benjamini-Hochberg post-hoc comparisons found that players who had the Dry game script liked Jo more than players who had the Original (Snark, Humor) game script (p < .0001) and Snark (No Humor) game script (p < .0001). Players who had the Humor (No Snark) script liked Jo more than players who had Original (Snark, Humor) game script (p < .0001) or the Snark (No Humor) game script (p < .0001).

Benjamini-Hochberg post-hoc comparisons also found that players who had the Dry game script found the game less humorous than players who had the Original (Snark, Humor) game script (p = 0.02) or the Humor (No Snark) game script (p = 0.003).

There were no significant effects between game script types in terms of how likely players were to complete the game, $\chi^2$ (df = 3) = 1.33, p = 0.72, their highest level reached, $F(3, 7112) = 0.01$, p = 0.998, or their amount of active time spent playing in the game, $\chi^2$ (df = 3) = 6.75, p = 0.08.

However, some relationships were found between player attitudes and these in-game measures. In specific, players who completed the game reported enjoying the game more, $\chi^2$ (df = 1) = 23.56, p < .0001, as well as finding the game more humorous, F(2, 7113) = 23.56, p = .0005], compared to students that had not completed the game. These findings match a similar finding from Dormann & Biddle (2006) and lends additional evidence for the benefits of humor in educational games to sustain engagement. However, player completion was not statistically significantly associated with how much they liked Jo, $\chi^2$ (df = 1) = 2.54, p = .111.

In terms of the highest level reached by a player, there were statistically significant but very small associations with player perceptions of the game. Spearman rank correlations showed a small but positive significant relationship between a player's enjoyment of the game and the highest game level they reached, $\rho$ = 0.08, t(7115) = 6.77, p < .0001, between finding a game humorous and the highest game level they reached, $\rho$ = 0.05, t(7115) = 4.05, p < .0001, and between liking Jo and the highest game level reached, t(7115) = 4.10, $\rho$ = 0.05, p < .0001.

In addition, players' reading level was statistically significantly associated with how many levels they played within the game, $\chi^2$ (df = 2) = 102.44, p < .0001. A higher reported reading level (Above Average > Average > Below Average) was associated with a higher maximum level reached by the player. Finally, players who reported higher reading levels also reported enjoying the game more, $\chi^2$ (df = 2) = 209.42, p < .0001. Originally, reading level was measured only to ensure that the changes in writing were perceived. Given these results, however, reading appears to be a prerequisite for succeeding in the game and enjoying the game fully.

## Conclusions

Using randomized experiments and embedded questioners with large anonymous audiences, we explored two elements of game writing, snark and humor. We found that humorous writing successfully led to players finding the game more humorous. Snarky writing's effect was more nuanced. While the snarky scripts did not produce a negative effect on enjoyment directly, it produced a less likable character. This was not the intent of the snarky writing, which was written in hopes that players would better identify with the player character and retain their own complex relationships with liking the topic of history. In general, both positive attitudes towards the game and reading level were associated with completing more of the game. Further work should be done to understand how game writing may produce a more relatable player character and exploring sub-populations in which this approach is most useful.

## Acknowledgements
This work was funded in part by NSF (DRL# 1907384/1907437) and the Wisconsin Department of Public Instruction.